\documentclass[a4paper,12pt]{article}
 \pdfoutput=1
\usepackage[pdftex]{graphicx}
\usepackage{subfigure}

\newcommand{\be}{\begin{equation}}
\newcommand{\ee}{\end{equation}}

\newcommand{\bea}{\begin{eqnarray}}
\newcommand{\eea}{\end{eqnarray}}

\newcommand{\p}{\partial}

\newcommand{\nn}{\nonumber \\}
\newcommand{\f}{\frac}

\newcommand{\ra}{\rightarrow}
\begin{document}
\thispagestyle{empty}
\begin{flushright}
{\bf arXiv:1507.06488}
\end{flushright}
\begin{center} \noindent \Large \bf  Entanglement Temperature With  Gauss-Bonnet  Term 
\end{center}

\bigskip\bigskip\bigskip
\vskip 0.5cm
\begin{center}
{ \normalsize \bf   Shesansu Sekhar Pal$^{a}$ \& Sudhakar Panda$^{b,\star}$\footnote{On leave from HRI, Allahabad}
}


\vskip 0.5 cm

${}^a$Department of Physics, Utkal University, Bhubaneswar 751004, India\\

${}^b$Institute of Physics, Sachivalaya Marg, Bhubaneswar 751005, India
\vskip 0.5 cm
\sf { shesansu${\frame{\shortstack{AT}}}$gmail.com \quad panda${\frame{\shortstack{AT}}}$iopb.res.in 
}
\end{center}
\centerline{\bf \small Abstract}

We compute the
entanglement temperature using the first law-like of thermodynamics, $\Delta E=T_{ent} \Delta S_{EE}$, 
up to  Gauss-Bonnet term in the 
Jacobson-Myers entropy functional in any arbitrary spacetime dimension. The computation is  done when the entangling region is the geometry  of a slab. We also show that such a Gauss-Bonnet term, which becomes a total derivative, when the co-dimension two hypersurface is   four dimensional, does not contribute to the finite term in the entanglement entropy. We observe   that the Weyl-squared term does not contribute to the entanglement entropy. It is important to note that the calculations are performed when the 
entangling region is very small and the energy is calculated using the normal Hamiltonian. 


\newpage
\section{Introduction and summary}

The recent study of entanglement entropy has drawn a lot of attention because of its remarkable similarity with the black hole entropy \cite{Srednicki:1993im}. The entanglement entropy is defined as von Neumann entropy: $S=-{\rm Trace}_A  \left(\rho_A Log~\rho_A\right)$, where $\rho_A$ is the  the reduced density matrix. For a spherical surface of radius, ${\cal R}$, the reduced desnity matrix is defined by tracing out the degrees of freedom that sits inside this radius, which is the complement of A. In which case, the von Neumann entropy depends on the area of the spherical region, $S\sim {\cal R}^2$. This von Neumann entropy is interpreted as the entropy seen by an observer sitting outside the radius  ${\cal R}$.

In the context of gauge-gravity duality \cite{Maldacena:1997re}, a prescription, strictly speaking a conjecture,  is suggested by Ryu and Takayanagi (RT) \cite{Ryu:2006bv} to calculate the von Neumann entropy in the gravitational side. This von Neumann entropy is called as entanglement entropy. The RT prescription suggests to consider a co-dimension two spatial hypersurface in  a  way such that its boundary coincides with the boundary of the region that we are interested in and then find out the area of the minimal surface.
Finally, the entanglement entropy is the ratio between  the area and $4G_N$, where $G_N$ is the Newton\rq{}s constant, (for  recent reviews, see \cite{Nishioka:2009un,Takayanagi:2012kg, Headrick:2013zda}). A claim of the proof of the RT conjecture is provided in different spacetime dimensions and with different entangling region  in  \cite{Casini:2011kv, Lewkowycz:2013nqa}. The subadditivity nature of the entanglement entropy is shown in \cite{Headrick:2007km}.

In this paper, we shall report the result of a holographic  calculation  of the entanglement entropy of a given region  upon inclusion of the terms up to  Gauss-Bonnet  term  in  the Jacobson-Myers (JM) entropy functional \cite{Jacobson:1993xs} in any arbitrary speactime dimension. The shape of the entangling region that we are interested in is that of the slab type. To recall, the prescription that we shall follow to carry out such a  calculation  is that given by Ryu and Takayanagi (RT) \cite{Ryu:2006bv}, of course without the higher derivative term.  


The JM entanglement entropy functional up to four derivative term is 
\be\label{jm}
4G_N S_{EE}=\int d^{d-1}\sigma \sqrt{det(g_{ab})}\left [1+\lambda_1 R(g)+\Lambda\left(R^2(g)-4R_{ab}(g)R^{ab}(g)+R_{abcd}(g)R^{abcd}(g) \right)\right],
\ee
where $\lambda_1$ and $\Lambda$ are unknown real coupling constants and are dimension full. 
A derivation  of the JM functional starting from the Einstein-Hilbert action with the higher derivative terms  are given in \cite{Dong:2013qoa, Camps:2013zua, Miao:2014nxa}. For our purpose, we do not require the full derivation of it.

The induced metric on the co-dimension two hypersurface is denoted as $g_{ab}=\p_aX^M\p_b X^N G_{MN}$, where $G_{MN}$ is the $d+1$ dimensional bulk spacetime geometry    and the hypersurface is described by $X^M$.  
The precise form of the hypersurface that follows  \cite{Pal:2013fha, Erdmenger:2014tba}
\bea\label{GB_eom}
&&   {\cal K}^S+\lambda_1\left(R{\cal K}^S-2R^{ab} {\cal K}^S_{ab}\right)+\Lambda\Bigg[ {\cal K}^S \left(R^2-4R_{ab}R^{ab}+ R_{a_1b_1c_1d_1} R^{a_1b_1c_1d_1}\right) -\nn&&4RR^{ab} {\cal K}^S_{ab}
+8R^{acbd}R_{cd} {\cal K}^S_{ab}-4R^{aecd}{R^b}_{ecd} {\cal K}^S_{ab}+8{R^a}_cR^{bc} {\cal K}^S_{ab}\Bigg]=0,
    \eea 
  which is essentially the equation of motion associated to the field $X^S$ and   $ {\cal K}^S\equiv  g^{ab}{\cal K}^S_{ab}$, whose precise
form ${\cal K}^S_{ab}=\p_a\p_bX^S-\gamma^c_{ab}\p_cX^S+\p_aX^M\p_bX^N\Gamma^S_{MN},$ 
where $\gamma^c_{ab}$ and $\Gamma^S_{MN}$ are the affine connections defined using the induced metric $g_{ab}$ and the bulk geometry, $G_{MN}$, respectively.  This particular form of the hypersurface holds good irrespective of the shape and size of the entangling region. 

 Let us note that without the higher derivative terms the equation of the hypersurface is derived earlier  in \cite{Hubeny:2007xt} and is called as the  extremal surface.  In what follows, we shall be interested in the strip type entangling region only.

In this paper, we shall compute  the correction to the  expression of the RT entanglement entropy by considering such higher derivative terms in the JM functional. This essentially means we are computing the entanglement entropy in the  finite \lq{}t Hooft limit\footnote{It simply follows from the AdS/CFT correspondence \cite{Maldacena:1997re}. The quantities $\lambda_1$ and $\Lambda$ are dimension full and the only allowed quantity that come is the size of the AdS spacetime, $R$. By the above mentioned duality $R\sim \ell_s \lambda^{1/4}$,  where $\ell_s$ and $\lambda$ are the string length and the 'tHooft coupling, respectively. It means $\lambda_1\sim \ell^2_s \lambda^{1/2}$ and 
 $\Lambda\sim \ell^4_s \lambda$.}. 
In fact, we shall be doing the calculation to the linear order in the couplings $\lambda_1$ and $\Lambda$, for simplicity. The result reads for the bulk geometry as AdS spacetime with radius\footnote{The size  $R_0$  is the solution to the bulk equation of motion with higher derivative term whereas $R$ is without \cite{Hung:2011xb}. The relationship is as follows: $R_0=\f{R}{\sqrt{f_{\infty}}}$, where $f_{\infty}$ is the positive real root of the  following cubic equation $1-f_{\infty}+\f{(d-2)(d-3)\lambda_1}{2R^2}f^2_{\infty}-\f{(d-2)(d-3)(d-4)(d-5)\Lambda}{3R^4}f^3_{\infty}=0$. In what follows, we shall express everything in terms of $R_0$ only, for simplicity. } $R_0$ as
\bea\label{ee_ell}
2 G_N S_{EE}&=&\f{L^{d-2}R^{d-1}_0}{(d-2)\epsilon^{d-2}}\left(1-\f{(d-1)(d-2)}{R^2_0}\lambda_1+\f{(d-1)(d-2)(d-3)(d-4)}{R^4_0}\Lambda \right)-\nn&& L^{d-2}R^{d-1}_02^{d-2} \pi^{\f{d-1}{2}}\f{\ell^{2-d}}{d-2}\left(\f{\Gamma\left(\f{d}{2(d-1)}\right)}{\Gamma\left(\f{1}{2(d-1)}\right)}\right)^{d-1}\times\nn &&\left(1+\f{(d-1)(d-2)(d-3)}{R^2_0}\lambda_1-\f{(d-1)^2(d-2)(d-3)(d-4)(d-5)}{(2d-1)R^4_0}\Lambda \right),
\eea
where the entangling region is taken as slab type. The slab that we are interested in is  $(d-1)$ dimensional. Along one direction, it is extended from $0$ to size $\ell$ and along the rest of the directions from $-L/2$ to $L/2$. It  is more properly  defined  in eq(\ref{def_strip}). The UV regulator is defined as $\epsilon$.

As found by RT and \cite{deBoer:2011wk}, the  entanglement entropy  has 
a divergent piece, which comes from UV and a  finite piece from IR. However, there are few salient features that are worth mentioning:

 a)  This result of the entanglement entropy makes sense only for $d\geq 3$.   For $d=3$, it is easy to notice  that the finite piece  does not receive any correction at finite \lq{}t Hooft coupling. However, the divergent piece can receive corrections.

b) The power of the  UV regulator, $\epsilon$, is independent of the value  of the \lq{}t Hooft coupling, i.e., it remains the same for infinite  as well as finite value of the \lq{}t Hooft coupling. However, the coefficient of it depends very well on the value of the \lq{}t Hooft coupling.

 c) The entanglement entropy depends on the quantity $\ell$ via power law type and this behavior is not changed even in the finite value of the 't Hooft coupling. However, $\ell$ does  depend  of the value  of the \lq{}t Hooft coupling through the turning point $r_{\star}$ to be followed latter.

  d) The  power of $L$, the size of the slab along the other spatial directions,  remains same both in the divergent as well as in the finite piece. This   is because, it does not play any  role in the determination of the extremal surface and comes as an over all factor in the integration of the entanglement entropy.  This is simply because of the translational invariance along these directions.

e) Note, for $d=5$, the Gauss-Bonnet term becomes topological, which means it won't contribute in  the determination of the hypersurface, $X^M(\sigma^a)$. This fact  is reflected in eq(\ref{sol}), to be followed. Moreover, such a term becomes a total derivative and does not contribute to the finite term in the entanglement entropy.

Doing a small perturbation  around the AdS spacetime geometry described by a parameter, $m$,   is shown to respect the first law-like of thermodynamics  $\Delta E=T_{ent} \Delta S_{EE}$ \cite{Bhattacharya:2012mi}, where $\Delta E$ corresponds to the energy of the excited state and $\Delta S_{EE}$ is the entanglement entropy. The entangling temperature depends on the size of the entangling region $\ell$.
  More importantly, this is  shown   in the  limit of small entangling region, $m \ell^d\ll 1$ as in  \cite{Bhattacharya:2012mi}. Some other interesting studies on the first law-like thermodynamics
  is done in \cite{Allahbakhshi:2013rda, Guo:2013aca, He:2013rsa,  Pang:2013lpa, Wong:2013gua, Caputa:2013lfa, Caputa:2013eka}.

In this paper,  we shall,  test the first law-like of thermodynamics, which reads as  $\Delta E=T_{ent} \Delta S_{EE}$, for the low-lying excited states   with higher derivative corrections in the entanglement entropy functional for any arbitrary spacetime dimensions.  
For the slab type entangling region, the entangling temperature takes the similar  form as before: $T_{ent}={\bf c}/\ell$, where $c$ is a constant and depends on the couplings. With the higher derivative term as written in the JM functional
\bea\label{exp_c}
{\bf c}&=&\f{2(d^2-1)}{\sqrt{\pi}}\left(1+\f{2(d-2)(d-3)}{R^2_0}\lambda_1-\f{2(d-2)(d-3)(d-4)(d-5)(3d^2-6d+2)}{(2d-1)(3d-1)R^4_0}\Lambda \right)\times\nn&&\left(\f{\Gamma\left(\f{d}{2(d-1)}\right)}{\Gamma\left(\f{1}{2(d-1)}\right)}\right)^2 \f{\Gamma\left(\f{d+1}{2(d-1)}\right)}{\Gamma\left(\f{1}{d-1}\right)}.
\eea
 
For $d=3$, the quantity $c$ is  universal in the sense that it is a pure number and 
 does not receive any corrections in the finite 'tHooft couplings.

The paper is organized as follows: In section 2, we give the  computational details of the entanglement entropy with higher derivative terms, up to Gauss-Bonnet term. In section 3, we show the first law-like  of thermodynamics with higher derivative terms and compute the entanglement temperature. In section 4, we re-visited the concavity and the specific heat and then conclude in section 5.
   
\section{Entanglement Entropy}       

Let us consider the  form of the entanglement entropy functional as written in eq(\ref{jm}) for the 
strip type entangling region. In this case the extremal hypersurface that follows is given in eq(\ref{GB_eom}) with
\be
{\cal K}^S_{ab}=\p_a\p_bX^S-\gamma^c_{ab}\p_cX^S+\p_aX^M\p_bX^N\Gamma^S_{MN},
 \ee 
where $\gamma^c_{ab}$ and $\Gamma^S_{MN}$ are the affine connections defined using the induced metric $g_{ab}$ and the bulk geometry, $G_{MN}$, respectively. The indices $a~b,~c$ etc run over the codimension two hypersurface, whereas $M,~N,~S$ etc run over the entire spacetime. 

\subsection{Slab/strip type entangling region}

Let us consider the following  background geometry possessing the translational symmetry along the temporal and spatial directions along with the rotational symmetry, with  diagonal form as
\be
ds^2_{d+1}=G_{MN}dx^Mdx^N=-g_{tt}(r)dt^2+g_{xx}(r)(dx^2_1+\cdots+dx^2_{d-1})+g_{rr}(r)dr^2.
\ee
The induced geometry that follows on a codimension two hypersurface is
\bea
&&X^t=0,\quad X^a=\sigma^a=x^a,\quad X^r=r(x_1),\quad r'\equiv \f{dr}{dx_1}\nn
&&ds^2_{d-1}=g_{ab}d\sigma^ad\sigma^b=\left[g_{xx}(r)+g_{rr}(r)\left(\f{dr}{dx_1}\right)^2 \right]dx^2_1+g_{xx}(r)(dx^2_2+\cdots+dx^2_{d-1}).
 \eea 

In order to carry out the explicit computations, we shall 
use the following form of the slab type entangling region
\be\label{def_strip}
0\leq x_1 \leq \ell,\quad -L/2\leq (x_2,\cdots,x_{d-1})\leq L/2.
 \ee

Various coordinate invariant quantities involving the Riemann tensor on the codimension two hypersurface  take the following from  
\bea
R&=&\f{(d-2)}{4g^2_{xx}(g_{xx}+g_{rr}r\rq{}^2)^2}\Bigg[2r\rq{}^4g_{xx}g\rq{}_{xx}g\rq{}_{rr}-(d-7)r\rq{}^2g_{xx}g\rq{}^2_{xx}-
 4r\rq{}\rq{}g^2_{xx}g\rq{}_{xx}-4r\rq{}^2g^2_{xx}g\rq{}\rq{}_{xx}\nn&&-4r\rq{}^4g_{xx}g_{rr}g\rq{}\rq{}_{xx}-
 (d-5)r\rq{}^4g_{rr}g\rq{}^2_{xx} \Bigg],\nn
 R_{ab}R^{ab}&=&\f{(d-2)}{16g^4_{xx}(g_{xx}+g_{rr}r\rq{}^2)^4}\Bigg[(d-2)\bigg(g_{rr}g\rq{}^2_{xx}r\rq{}^4+g_{xx}(2g\rq{}^2_{xx}r\rq{}^2+g\rq{}_{rr}g\rq{}_{xx}r\rq{}^4-2g_{rr}r\rq{}^4g\rq{}\rq{}_{xx})-\nn&&2g^2_{xx}(r\rq{}^2g\rq{}\rq{}_{xx}+g\rq{}_{xx}r\rq{}\rq{})\bigg)^2+\bigg((d-4) g_{rr}g\rq{}^2_{xx}r\rq{}^4+2g^2_{xx}(r\rq{}^2 g\rq{}\rq{}_{xx}+g\rq{}_{xx}r\rq{}\rq{})+\nn && g_{xx} r\rq{}^2\bigg[(d-5)g\rq{}^2_{xx}-g\rq{}_{rr}g\rq{}_{xx}r\rq{}^2+2g_{rr}r\rq{}^2g\rq{}\rq{}_{xx}\bigg]\bigg)^2\Bigg],\nn
 R_{abcd}R^{abcd}&=&\f{(d-2)}{8g^4_{xx}(g_{xx}+g_{rr}r\rq{}^2)^4}\Bigg[(d-3)g\rq{}^4_{xx}r\rq{}^4(g_{xx}+g_{rr}r\rq{}^2)^2+
 2\bigg(g_{rr}g\rq{}^2_{xx}r\rq{}^4+\nn&&g_{xx}(2g\rq{}^2_{xx}r\rq{}^2+g\rq{}_{rr}g\rq{}_{xx}r\rq{}^4-2g_{rr}r\rq{}^4g\rq{}\rq{}_{xx}) -2g^2_{xx}(r\rq{}^2g\rq{}\rq{}_{xx}+g\rq{}_{xx}r\rq{}\rq{})\bigg)^2\Bigg].
\eea

Using these informations, we can compute $GB=R_{abcd}R^{abcd}-4R^{ab}R_{ab}+R^2$ and comes as
\bea
GB&=&\f{(d-2)(d-3)(d-4)}{16g^4_{xx}(g_{xx}+g_{rr}r\rq{}^2)^3}g\rq{}^2_{xx}r\rq{}^2\Bigg[(d-9)g_{rr}g\rq{}^2_{xx}r\rq{}^4+g_{xx}r\rq{}^2\bigg((d-13)g\rq{}^2_{xx}-
4g\rq{}_{rr}g\rq{}_{xx}r\rq{}^2+\nn&&8g_{rr}r\rq{}^2g\rq{}\rq{}_{xx} \bigg)+8g^2_{xx}(r\rq{}^2g\rq{}\rq{}_{xx}+g\rq{}_{xx}r\rq{}\rq{})\bigg]
\eea

In passing we must mention that upon doing the calculation of the  Weyl-squared term defined as: $Weyl^2\equiv R_{abcd}R^{abcd}-\f{4}{d-3}R^{ab}R_{ab}+\f{2}{(d-2)(d-3)}R^2$ vanishes identically for the strip type of entangling region. Hence, to conclude the Weyl-squared term does not contribute anything to the entanglement entropy functional.

Substituting all these into the equations of motion eq(\ref{GB_eom}) and denoting $x'_1\equiv \f{dx_1}{dr}$  gives
\be
\f{d}{dr}\left(\f{g^{d/2}_{xx}x\rq{}_1}{\sqrt{g_{rr}+g_{xx}x\rq{}^2_1}}-\f{(d-2)(d-3)g^{\f{d-4}{2}}_{xx}x\rq{}_1g\rq{}^2_{xx}}{4(g_{rr}+g_{xx}x\rq{}^2_1)^{3/2}}\lambda_1+\f{(d-2)(d-3)(d-4)(d-5)g^{\f{d-8}{2}}_{xx}g\rq{}^4_{xx}x\rq{}_1\Lambda}{16(g_{rr}+g_{xx}x\rq{}^2_1)^{5/2}} \right)=0.
\ee

Solving the equation to leading order in the couplings
\be\label{sol}
x\rq{}_1(r)=\f{c\sqrt{g_{rr}}}{\sqrt{g^d_{xx}-c^2g_{xx}}}+\f{c(d-2)(d-3)g\rq{}^2_{xx}\lambda_1}{4g^2_{xx}\sqrt{g_{rr}}\sqrt{g^d_{xx}-c^2g_{xx}}}-\f{c(d-2)(d-3)(d-4)(d-5)g\rq{}^4_{xx}\sqrt{g^d_{xx}-c^2g_{xx}}}{16g^{3/2}_{rr}g^{d+4}_{xx}}\Lambda,
\ee
where $c$ is the constant of integration. It is chosen to take $c=g^{\f{d-1}{2}}_{xx}(r_{\star})$, because we have opted  the following boundary condition  at $r=r_{\star}$: the quantity ${\rm Limit}_{r\ra r_{\star}}x\rq{}_1=\infty$, diverges. 

Substituting it into the entanglement entropy functional  gives
\bea\label{general_ee}
2G_NS_{EE}&=&L^{d-2}\int dr\Bigg(\f{\sqrt{g_{rr}}g^{\f{2d-3}{2}}_{xx}}{\sqrt{g^{d-1}_{xx}-c^2}}-\f{(d-2)\lambda_1}{4g^{3/2}_{rr}g^{5/2}_{xx}\sqrt{g^{d-1}_{xx}-c^2}}\bigg(6c^2g_{rr}g\rq{}^2_{xx}+(d-5)g_{rr}g^{d-1}_{xx}g\rq{}^2_{xx}+\nn&&
2g_{xx}(c^2-g^{d-1}_{xx})(g\rq{}_{rr}g\rq{}_{xx}-2g_{rr}g\rq{}\rq{}_{xx})\bigg)+\Lambda\f{(d-2)(d-3)(d-4)g\rq{}^2_{xx}}{16g^{5/2}_{rr}g^{\f{2d+7}{2}}_{xx}}\sqrt{g^{d-1}_{xx}-c^2}\times \nn&&\bigg(g_{rr}g\rq{}^2_{xx}[2c^2(d+5)+(d-9)g^{d-1}_{xx}]+4g_{xx}(c^2-g^{d-1}_{xx})(g\rq{}_{rr}g\rq{}_{xx}-2g_{rr}g\rq{}\rq{}_{xx}) \bigg)\Bigg).
\eea

Let us evaluate the above integral by consider a specific example. For  our purpose,  we shall take the example of AdS spacetime.
For this purpose, we put the boundary to be at $r=0$
\be
ds^2_{d+1}=\f{R^2_0}{r^2}\left(-dt^2+dx^2_1+\cdots+dx^2_{d-1}+dr^2\right)
\ee

The result reads as
\bea\label{ee_rs}
&&2G_NS_{EE}= \f{L^{d-2}R^{d-1}_0}{(d-2)\epsilon^{d-2}}\left(1-\f{(d-1)(d-2)}{R^2_0}\lambda_1+\f{(d-1)(d-2)(d-3)(d-4)}{R^4_0}\Lambda \right)-\nn&&\f{\sqrt{\pi}L^{d-2}R^{d-1}_0}{(d-2)r^{d-2}_{\star}}\f{\Gamma\left( \f{d}{2(d-1)}\right)}{\Gamma\left( \f{1}{2(d-1)}\right)}\left(1+\f{(d-2)(d-3)}{R^2_0}\lambda_1-\f{(d-1)(d-2)(d-3)(d-4)(d-5)}{(2d-1)R^4_0} \Lambda\right),\nn
\eea
where  the integral over $r$ is  performed from the the UV cutoff, $r=\epsilon$,  to IR $r=r_{\star}$.  
Now we can use the following relationship between $\ell$ and $r_{\star}$ to leading order in the couplings
\be\label{l_star_m_zero}
\ell/2=r_{\star}\f{\sqrt{\pi}\Gamma\left( \f{d}{2(d-1)}\right)}{\Gamma\left(\f{1}{2(d-1)} \right)}\left(1-\f{(d-2)(d-3)}{R^2_0}\lambda_1+
\f{(d-1)(d-2)(d-3)(d-4)(d-5)}{(2d-1)R^2_0}\Lambda \right).
\ee
This follows from eq(\ref{sol}), upon doing the integration. On substituting this into the entanglement entropy as written in eq(\ref{ee_rs}) gives us eq(\ref{ee_ell}).

\subsection{ For $d=5$}

It is interesting to note that for a specific dimension, $d$, the Gauss-Bonnet term does not contribute anything to the equation of motion. This happens when $d=5$, in fact, in this case, the Gauss-Bonnet term in the entanglement entropy functional becomes a pure topological term. In fact, it can be expressed as a total derivative. The purpose of this subsection is to show that this   total derivative term does not contribute anything to the finite piece but does to the divergent piece of the entanglement entropy. 

For $d=5$, the Gauss-Bonnet term can be expressed as
\bea
&&\Lambda \int^{\ell}_0 dx_1 \sqrt{det(g_{ab})} \left(R^2(g)-4R_{ab}(g)R^{ab}(g)+R_{abcd}(g)R^{abcd}(g) \right)=
-\f{3}{2}\Lambda\times \nn&&\int^{r_{\star}}_{\epsilon} dr \f{g'^2_{xx}\left((g'^2_{xx}-2g_{xx}g''_{xx})(g_{rr}+g_{xx}x'^2_1) +g_{xx}g'_{xx}(g'_{rr}+g'_{xx}x'^2_1+2g_{xx}x'_1x''_1)\right)}{g^{\f{5}{2}}_{xx}(g_{rr}+g_{xx}x'^2_1)^{\f{5}{2}}}=\Lambda\times\nn
&& \int^{r_{\star}}_{\epsilon} dr \f{d}{dr}\left(\f{g'^3_{xx}}{g^{\f{3}{2}}_{xx}(g_{rr}+g_{xx}x'^2_1)^{\f{3}{2}}} \right)=\Lambda\left(\f{g'^3_{xx}}{g^{\f{3}{2}}_{xx}(g_{rr}+g_{xx}x'^2_1)^{\f{3}{2}}} \right)^{r_{\star}}_{\epsilon}.
\eea

Using the bulk as AdS spacetime structure and evaluating it gives the desired result as written above.
So to conclude, we find that inclusion of a total derivative term in the entanglement entropy functional does not contribute anything to the finite term of the 
entanglement entropy.

\subsection{Scaling symmetry}

Let us demand that under the following  scale transformation of the coordinates, the metric components transformations as
\be
x^{M}\ra \lambda x^M,\quad G_{MN}\ra \lambda^{-2} G_{MN}
\ee
for which the length between two points in the bulk geometry remains invariant. This  property holds for the AdS spacetime apart from other symmetries. On the codimension two hypersurface, we also want 

\be
\sigma^{a}\ra \lambda \sigma^a,\quad g_{ab}\ra \lambda^{-2} g_{ab},\quad g_{ab}d\sigma^ad\sigma^b\ra \lambda^0 g_{ab}d\sigma^ad\sigma^b.
\ee

Then it simply follows that the Lorentz scalar quantities made out of the  Ricci curvatures behaves as 
\be
R(g)\ra \lambda^0 R(g),\quad R_{ab}(g)R^{ab}(g)\ra \lambda^0R_{ab}(g)R^{ab}(g),\quad R_{abcd}(g)R^{abcd}(g)\ra \lambda^0R_{abcd}(g)R^{abcd}(g).
\ee

It is very easy to check that the JM entanglement entropy functional eq(\ref{jm}) does not scale under the above transformation. It is because of this 
scaling symmetry the size of the entangling region comes as an over all factor in the computation of the entanglement entropy functional.
One can notice that each term in the following expression
\be
1+\lambda_1 R(g)+\Lambda\left(R^2(g)-4R_{ab}(g)R^{ab}(g)+R_{abcd}(g)R^{abcd}(g) \right)
\ee
does not scale as said above. In fact, we expect it not to scale for a scale invariant bulk geometry. It is also expected that the size $L$ should not enter in the computation of this expression. Hence, it is quite natural to expect that the size of the entangling region should come as an over all factor. Note that both $L,~\ell$ scale as $(L,~\ell)\ra \lambda (L,~\ell)$. This means the turning point $r_{\star}$ also scale linearly in $\lambda$ by virtue of eq(\ref{l_star_m_zero}). It also means that the quantity $c$ scales as $c\ra \lambda^{-(d-1)} c$.

From our studies of thermodynamics, it is well known that if we scale the volume $V\ra \Lambda V$ and energy  $E\ra \Lambda E$, then the entropy scales as $S(V,~E)\ra S(\Lambda V, \Lambda E)=\Lambda S(V,~E)$. This is just the homogeneity of order one property obeyed by the entropy. 

Note for the AdS geometry the energy vanishes which means the only relevant function is $S(V)$. Recall, for the slab type entangling region, the volume $V\equiv \ell L^{d-2}$. In order to have the desired scaling for the volume, we must scale $(\ell,~L)\ra \Lambda^{\f{1}{d-2}} (\ell,~L)$. This means $\lambda=\Lambda^{\f{1}{d-2}}$. In which case, the finite part of the entanglement entropy does not scale but the same cannot be said about the singular part. As it is a bit ambiguous. 

\section{Small Fluctuations}
In this section, we shall check the first law-like of thermodynamics by considering  small fluctuations of the bulk geometry along the lines of \cite{Bhattacharya:2012mi} but now with the higher derivative terms. In order to check such a law, the entanglement entropy will play the role of the entropy and the change in energy will be the energy of the low lying excited states and then their ratio will give us the analogue of the temperature which will be called as entanglement temperature. 

For the following type of fluctuation of the AdS geometry 
\be
g_{tt}=\f{R^2_0}{r^2(1+mr^d)},\quad g_{xx}=\f{R^2_0}{r^2},\quad g_{rr}=\f{R^2_0}{r^2(1-mr^d)},
\ee
where $m$ is the fluctuating parameters. We can do the necessary calculations very easily using the expressions for  the 
 entanglement entropy as given in eq(\ref{general_ee}).
  In what follows, we shall assume that $m\ell^d \ll 1$ \cite{Bhattacharya:2012mi}. This condition can as well be re-written as $mr^d_{\star}\ll 1$, so as to keep terms to linear order in the parameter, $m$. Now, to satisfy such a requirement means we must take   $r_{\star}$ very close to UV because the boundary is at $r=0$.

Substituting these metric fluctuations into eq(\ref{general_ee}) and carrying out the necessary $r$ integrals from UV to IR results in
\bea
2G_NS_{EE}&=& \f{L^{d-2}R^{d-1}_0}{(d-2)\epsilon^{d-2}}\left(1-\f{(d-1)(d-2)}{R^2_0}\lambda_1+\f{(d-1)(d-2)(d-3)(d-4)}{R^4_0}\Lambda \right)-\nn \f{\sqrt{\pi}L^{d-2}R^{d-1}_0}{(d-2)r^{d-2}_{\star}}&&\f{\Gamma\left( \f{d}{2(d-1)}\right)}{\Gamma\left( \f{1}{2(d-1)}\right)}\left(1+\f{(d-2)(d-3)}{R^2_0}\lambda_1-\f{(d-1)(d-2)(d-3)(d-4)(d-5)}{(2d-1)R^4_0} \Lambda\right)\nn&+&
\f{m}{4} L^{d-2} r^2_{\star}R^{d-1}_0\sqrt{\pi}\f{\Gamma\left( \f{d}{(d-1)}\right)}{\Gamma\left( \f{d+1}{2(d-1)}\right)}\times\nn&&\left(1-\f{(d-2)(d-3)}{R^2_0}\lambda_1+\f{3(d-1)(d-2)(d-3)(d-4)(d-5)}{(3d-1)R^4_0}\Lambda \right)
\eea

Now, we want to express $r_{\star}$ in terms of $\ell$ to leading order in $m$ and it comes as
\bea
\f{\ell}{2}&=&r_{\star}\f{\sqrt{\pi}\Gamma\left( \f{d}{2(d-1)}\right)}{\Gamma\left(\f{1}{2(d-1)} \right)}\left(1-\f{(d-2)(d-3)}{R^2_0}\lambda_1+\f{(d-1)(d-2)(d-3)(d-4)(d-5)}{(2d-1)R^2_0}\Lambda \right)+\nn&&m
\f{\sqrt{\pi}r^{d+1}_{\star}\Gamma\left( \f{d}{(d-1)}\right)}{2(d+1)\Gamma\left(\f{d+1}{2(d-1)} \right)}\left(1-\f{(d-2)(d-3)}{R^2_0}\lambda_1+\f{3(d-1)(d-2)(d-3)(d-4)(d-5)}{(3d-1)R^2_0}\Lambda \right)\nn
\eea

Now, we can use such a relation to  re-express the entanglement entropy in terms of $\ell$ as
\bea\label{ee_ell_m}
2G_NS_{EE}(m)&=&\f{L^{d-2}R^{d-1}_0}{(d-2)\epsilon^{d-2}}\left(1-\f{(d-1)(d-2)}{R^2_0}\lambda_1+\f{(d-1)(d-2)(d-3)(d-4)}{R^4_0}\Lambda \right)-\nn&& L^{d-2}R^{d-1}_02^{d-2} \pi^{\f{d-1}{2}}\f{\ell^{2-d}}{d-2}\left(\f{\Gamma\left(\f{d}{2(d-1)}\right)}{\Gamma\left(\f{1}{2(d-1)}\right)}\right)^{d-1}\times\nn &&\left(1+\f{(d-1)(d-2)(d-3)}{R^2_0}\lambda_1-\f{(d-1)^2(d-2)(d-3)(d-4)(d-5)}{(2d-1)R^4_0}\Lambda \right)+\nn&&
m\f{(d-1)L^{d-2}R^{d-1}_0\ell^2}{16\sqrt{\pi}(d+1)}\left(\f{\Gamma\left(\f{1}{2(d-1)}\right)}{\Gamma\left(\f{d}{2(d-1)}\right)}\right)^2\f{\Gamma\left(\f{d}{(d-1)}\right)}{\Gamma\left(\f{d+1}{2(d-1)}\right)}\times\nn&&
\left(1-\f{3(d-2)(d-3)}{R^2_0}\lambda_1+\f{(d-1)(d-2)(d-3)(d-4)(d-5)(12d-5)}{(2d-1)(3d-1)R^4_0}\Lambda \right)\nn
 \eea 

There follows the expression of the change in entanglement entropy:  $\Delta S_{EE}\equiv S_{EE}(m,~\ell)-S_{EE}(m=0,~\ell)$, which is finite and does not depend on the UV cutoff, $\epsilon$.

\paragraph{Energy:} The energy of such a perturbed geometry can be calculated using the expression of the energy momentum tensor as given in \cite{Dehghani:2006dh} with  the prescription given in \cite{Balasubramanian:1999re}.

\be
\Delta M=\int d^{d-1}x \sqrt{det(\sigma_{ij})}~ N~u^M u^N T_{MN},
\ee
\be
T_{MN}=\f{1}{8\pi G_N}\left(K_{MN}-KG_{MN} +2\lambda (3 J_{MN}-J G_{MN})+3{\tilde\Lambda}(5P_{MN}-PG_{MN})+\f{(d-1)}{{\tilde R}}G_{MN}\right),
\ee
where $K$ is the trace of the extrinsic curvature. The other quantities are defined as
\bea
J_{MN}&=&\f{1}{3}\left(2KK_{ML}K^L_N+K_{LS}K^{LS}K_{MN}-2K_{ML}K^{LS}K_{SN}-K^2K_{MN} \right),\nn
P_{MN}& =&
\f{1}{5}\Bigg[\left(K^4 - 6K^2K^{LS}K_{LS} + 8KK_{LS}K^S_PK^{PL} - 6K_{LS}K^{SP}K_{PR}K^{RL} + 3(K^{LP}K_{LP})^2\right)\nn K_{MN}
&-&(4K^3 -12KK_{LS}K^{LS} + 8K_{LS}K^S_PK^{PL})K_{MR}K^R_N - 24KK_{MS}K^{SP}K_{PR}K^R_N\nn
&+&(12K^2 -12K_{LP}K^{LP} )K_{MS}K^{SR}K_{RN} + 24K_{ML}K^{LP}K_{PS}K^{SR}K_{RN}\Bigg]
\eea
The quantity, ${\tilde R}$, acts as a regulator and 
we can expand  ${\tilde R}=R_0+\lambda R_1+{\tilde\Lambda} R_2$ to linear order in the couplings.
The sizes $R_1$ and $R_2$ will be determined by demanding that $T_{tt}$ becomes finite as we approach the boundary. Or in the limit of  $m\ra 0$, the  $T_{tt}$ component should vanish as well \cite{Balasubramanian:1999re}. The quantities $R_{1,2}$ are
\be
R_1=\f{2}{3R_0}(d-2)(d-3),\quad R_2=-\f{3}{5R^3_0}(d-2)(d-3)(d-4)(d-5), 
\ee
where $R_0$ is related to $R$ as $R_0=R/\sqrt{f_{\infty}}$ and $f_{\infty}$ obeys the following equation: $1-f_{\infty}+\f{(d-2)(d-3)}{2R^2}\lambda_1 f^2_{\infty}-\f{(d-2)(d-3)(d-4)(d-5)}{3R^4}\Lambda f^3_{\infty}=0$ \cite{Hung:2011xb}. In fact, we shall not express any of our results in terms of $R$, for simplicity. 

which in our case using $N=\sqrt{g_{tt}},~u^t=1/\sqrt{g_{tt}}$ and $ \sqrt{det(\sigma)_{ij}}=g^{\f{d-1}{2}}_{xx}$, gives
\be\label{exp_mass}
\Delta M=\int d^{d-1}x~ g^{\f{d-1}{2}}_{xx}~\f{T_{tt}}{\sqrt{g_{tt}}}
\ee

The expression of the various components of the extrinsic curvatures are $K_{tt}=-\f{ g\rq{}_{tt}}{2\sqrt{g_{rr}}},~ K_{xx} =\f{ g\rq{}_{xx}}{2\sqrt{g_{rr}}}$ and $K=\f{ g\rq{}_{tt}}{2g_{tt}\sqrt{g_{rr}}}+\f{(d-1) g\rq{}_{xx}}{2g_{xx}\sqrt{g_{rr}}}$.
Substituting all these along with the AdS spacetime into the above integral results in 

\be
E(m)=\f{m(d-1)\ell}{16 \pi G_N}L^{d-2}R^{d-1}_0\left(1-\f{2(d-2)(d-3)}{R^2_0}\lambda+\f{3(d-2)(d-3)(d-4)(d-5)}{R^4_0}{\tilde\Lambda} \right).
\ee
The bulk couplings $\lambda,~{\tilde\Lambda}$ are related to the couplings appear in the entanglement entropy functional are
$\lambda=\f{\lambda_1}{2}$ and ${\tilde\Lambda}=\f{\Lambda}{3}$ \cite{Hung:2011xb}. It is easy to see that for vanishing $m$, the energy vanishes, which is true for the AdS spacetime \cite{Balasubramanian:1999re}. Hence $\Delta E=E(m)$.

Having got all the desired expression for the change in the entanglement entropy and the change in energy means, we can evaluate the entanglement temperature, $T_{ent}=\Delta E/\Delta S_{EE}\equiv \f{{\bf c}}{\ell}$. Upon doing the calculations, we obtained eq(\ref{exp_c}), which is true for any $d\geq 3$ when the entangling region is of the slab type. One of the interesting point is that the quantity ${\bf c}$ is independent of the parameter $m$, even though the energy and the entanglement entropy do depend on it.

There exists a difference in the way the entanglement temperature is calculated in comparison to that in the  black hole physics, even though there exists a first law-like relation of thermodynamics in both cases. Note that the  entanglement temperature is not the inverse periodicity associated to any shrinking one-cycle as in black hole physics, because there does not exist any on the co-dimension two hypersurface. In fact, it takes the following form
\bea
ds^2_{d-1}&=&dx^2_1\Bigg(\f{g_{xx}^d}{c^2}+\f{(d-2)(d-3)g\rq{}^2_{xx}(c^2-g^{d-1}_{xx})}{2c^2 g_{rr}g_{xx}}\lambda_1+\nn&&\f{(d-2)(d-3)(d-4)(d-5)g\rq{}^4_{xx}(g^{d-1}_{xx}-c^2)^2}{8c^2 g^{2+d}_{xx} g^2_{rr}}\Lambda\Bigg)+g_{xx}(dx^2_2+\cdots+dx^2_{d-1}),
\eea
where $c^2=g^{d-1}_{xx}(r_{\star})$. Note that for the AdS type 
background geometry the co-dimension two geometry   is not conformal to flat space for any dimension $d\geq 3$. Even though, it is not conformal to flat space but it respects the scaling symmetry as discussed previously for $m=0$.

\section{ Concavity \& Specific heat }

\paragraph{Concavity:} It is suggested  that the entanglement  
entropy obeys the  strong sub-additivity property and this  followed by looking at 
 the concavity of entanglement entropy \cite{Bhattacharya:2012mi}. It means the concavity automatically implies the sub-additivity of entanglement entropy. The property concavity  suggests the second derivative of the entanglement entropy with respect to the size of the system should be negative: $\f{d^2S_{EE}}{d\ell^2}\leq 0$. 

In our case,  the entanglement entropy at finite 'tHooft coupling  takes the following form  as in eq(\ref{ee_ell_m}),
\be\label{form_s_ee}
S_{EE}=S_{\epsilon}-\f{\alpha}{G_N \ell^{d-2}} L^{d-2} R^{d-1}_0+\f{m\beta}{G_N} L^{d-2} R^{d-1}_0\ell^2,
\ee
where $S_{\epsilon}$ is the UV regulator dependent part and the value of $\alpha,~\beta$ can be obtained by comparing with  eq(\ref{ee_ell_m}), which are essentially positive. Demanding the concavity condition gives
\be
\f{d^2S_{EE}}{d\ell^2}\leq 0\Longrightarrow ~\beta \leq \f{\alpha(d-1)(d-2)}{2m\ell^d}.
\ee
Remember, we work in a limit for which $m\ell^d\ll 1$, and this means for finite $\alpha$ this condition is fulfilled automatically. Hence, according to \cite{Bhattacharya:2012mi} the entanglement entropy at finite 'tHooft coupling obeys  the strong sub-additivity property.

\paragraph{Specific heat:} It is suggested in \cite{Bhattacharya:2012mi} that the specific heat that follows from the calculation of the entanglement entropy becomes positive\footnote{We feel that such a calculation should be revisited.}.   For unit dynamical exponent, the temperature goes as $T_{ent}={\bf c}/\ell$. 
In which case, the  specific heat defined as 
\bea
C\equiv T_{ent}\f{\p (\Delta S_{EE})}{\p T_{ent}}&=&
-2\f{m\beta}{G_N}\f{{\bf c}VR^{d-1}_0}{T_{ent}},
\eea
where $V=\ell L^{d-2}$ is the volume of entangling region under study and $\Delta S_{EE}$ is the last term in eq(\ref{form_s_ee}).


There are a couple of comments in order:

(a) The volume is a function of temperature, $V=\f{{\bf c} L^{d-2}}{T_{ent}}$ and it goes as inverse of the entanglement temperature. Form which it follows trivially that $\f{T_{ent}}{V}\f{dT_{ent}}{dV}=-1$. 

(b)The negative form of the heat capacity (because $\beta >0$) implies the system under study is unstable, which is in agreement with the fact that we are trying to understand the properties of an excited state. Just to compare, the  temperature dependence of the heat capacity for a black hole with spatial horizon in  $d+1$ dimensional  AdS spacetime, $C\sim T^{d-1}_H$, where $T_H$ is the Hawking temperature.

(c) In thermodynamics, upon scaling the volume $V\ra \Lambda V$ and energy $E\ra \Lambda E$, makes the entropy to scale as $S\ra \Lambda S$. This is due to the  homogeneity condition of the entropy. However, such a simple relation does not hold in the case of entanglement entropy. The change in entanglement entropy,  $\Delta S_{EE}$ goes as $\Delta S_{EE}\ra \Lambda^{\f{d}{d-1}} \Delta S_{EE}$.

Generically, the holding  of the  second law of thermodynamics, $\Delta S_{EE}\geq 0$,  suggests the positivity of the specific heat. However, in the present case, it looks like that the excited states does not obey the second law of thermodynamics. 

\section{Conclusion}

To conclude, we have calculated the contribution of the   higher derivative terms up to Gauss-Bonnet term in the JM entanglement entropy functional by considering the background spacetime as AdS. To carry out the calculations  we have considered  the entangling region is of the slab type. At the end,  the form of the finite piece of the  entanglement entropy up to a sign    goes as $S_{EE}(finite)\sim \f{L^{d-2}R^{d-1}_0\ell^{2-d}}{G_N} f(\lambda_1,~\Lambda)$, where the functional form of the couplings $ f(\lambda_1,~\Lambda)$ is very difficult to predict and follows by doing the calculations. The dependence on $L,~R_0$ and $\ell$ follows from translational invariance and dimensional analysis. However, the absence of the information related to the entangling region like $L$ and $\ell$ in  $ f(\lambda_1,~\Lambda)$ follows simply  from the scale invariance.  Moreover, we observed that  the Weyl squared term does not contribute anything to the entanglement entropy functional.

For $d=5$, i.e., when the bulk spacetime is $5+1$ dimensional, it is easy to see that  the Gauss-Bonnet term in the JM entropy functional   becomes a pure topological term. In fact, this piece becomes a total derivative term. Moreover, upon calculating we find  that it does not contribute to the finite term but   does contribute to the UV regulator dependent term  in the entanglement entropy.

Recently, it was suggested in \cite{Astaneh:2014sma} by adding a different kind  of exotic term to the entanglement entropy functional, which is a total derivative term in the bulk,   and was shown that such a term does contribute to the finite part of the entanglement entropy. So, the future question of interest would be to find under what condition does a total derivative term contributes to the entanglement entropy, generically?

After doing a small perturbation described by a parameter, $m$, around the AdS spaetime, we find that the change in entanglement entropy up to an overall couplings goes as $\Delta S_{EE}(m,\ell)\equiv S_{EE}(m,\ell)-S_{EE}(m=0,\ell)\sim \f{mL^{d-2}R^{d-1}_0\ell^2}{G_N}$. The energy associated to such a geometry up to an overall couplings goes as $\Delta E\sim \f{mL^{d-2}R^{d-1}_0\ell}{G_N}$. Now, if we 
we  demand that there exists a first law-like of thermodynamics as in \cite{Bhattacharya:2012mi} then the entanglement temperature defined as $T_{ent}=\f{\Delta E}{\Delta S_{EE}}=\f{{\bf c}}{\ell}$, where the quantity ${\bf c}$ depends on the couplings and the dimension of the spacetime.

In this paper, we have worked in the limit when the entangling region is very small, $m\ell^d \ll 1$,  and checked the first law-like of thermodynamics using the normal Hamiltonian for the slab type geometry. However, in  \cite{Blanco:2013joa} and \cite{Faulkner:2013ica}, the first law-like of thermodynamics is checked using the modular Hamiltonian for the ball-shaped entangling region. 
It is certainly very interesting to check this for other type of entangling  geometries,   which we leave for future studies.

\paragraph{Acknowledgment:} We are happy to give thanks to the organizers of ISM 2014 held at Puri for providing an enjoyable atmosphere, where a part of the work is done. Many thanks to the anonymous referee for bringing  our attentions to 
two important references.

\end{document}